\documentclass[12pt]{article}

\usepackage[utf8]{inputenc}
\usepackage[T1]{fontenc}
\usepackage{amsmath,amssymb,amsfonts,mathtools}
\usepackage{bm}
\usepackage{booktabs}
\usepackage{geometry}
\usepackage{graphicx}
\usepackage{microtype}
\usepackage{enumitem}
\usepackage{pgfplots}
\usepackage{tikz}
\usepackage{xcolor}
\usepackage{hyperref}
\usepackage[nameinlink,noabbrev]{cleveref}
\pgfplotsset{compat=1.18}

\hypersetup{
  colorlinks=true,
  linkcolor=blue,
  citecolor=blue,
  urlcolor=blue
}

\newcommand{\Qsc}{\mathbb{Q}}
\newcommand{\dd}{\mathrm{d}}
\newcommand{\Ord}{\mathcal{O}}
\newcommand{\Tsg}{T_{\kappa}}
\newcommand{\Tfl}{T_{\mathrm{FL}}}
\newcommand{\Scorr}{S_{\mathrm{corr}}}
\newcommand{\LCnabla}{\mathring{\nabla}}
\newcommand{\LCR}{\mathring{R}}
\newcommand{\LCG}{\mathring{G}}
\newcommand{\Lag}{\mathcal{L}}
\newcommand{\cF}{\mathcal{F}}

\title{\textbf{Thermodynamic Consistency of Logarithmic Entropy Corrections on the Schwarzschild Branch of \texorpdfstring{$f(\Qsc)$}{f(Q)} Gravity and a Superradiance No-Go Result}}
\author{Wen-Xiang Chen\\
School of Electronic Information\\
Guangzhou City University of Technology\\
\texttt{wxchen4277@qq.com}}
\date{}

\begin{document}
\maketitle

\begin{abstract}
We give a fully explicit derivation of logarithmically corrected thermodynamics and scalar scattering on the Schwarzschild branch of symmetric teleparallel $f(\Qsc)$ gravity. The metric and the flat, torsion-free affine connection are treated as independent variables. We define the nonmetricity tensor, its traces, the disformation, the superpotential, and the scalar $\Qsc$; display the metric and connection equations; and show in detail why the linear choice $f(\Qsc)=\Qsc$ is dynamically equivalent to general relativity up to a boundary term. Solving the vacuum equations step by step yields the Schwarzschild metric and the mass--horizon relation $M=r_+/(2G)$.

The leading entropy is justified on this branch by the Noether-charge result and independently reconstructed from the classical first law. We then apply the phenomenological correction $S_{\mathrm{corr}}=S_0+\alpha\ln(S_0/S_\star)$ and explicitly distinguish the surface-gravity temperature, the first-law effective temperature, and a hypothetical backreacted Hawking temperature. Keeping the geometry and ADM mass fixed gives
\[
T_{\mathrm{FL}}=\frac{r_+}{4(\pi r_+^2+\alpha G)},
\]
whereas the geometric Hawking temperature remains $T_\kappa=1/(4\pi r_+)$. The corresponding heat capacities and Helmholtz diagnostic are derived without omitted algebra. The formal heat-capacity pole lies at $S_0=\alpha$, precisely where the first-order logarithmic expansion is uncontrolled; it is therefore not evidence for a physical phase transition.

Finally, beginning with the four-dimensional Klein--Gordon equation, we derive the tortoise-coordinate radial equation, its real effective potential, the conserved Wronskian, and the reflection--transmission relation. For a neutral scalar on a static neutral Schwarzschild background, $|\mathcal R|^2\leq1$. Since the entropy correction neither changes the metric nor creates a horizon chemical potential, it produces no first-order shift of a superradiant frequency window or amplification factor. A nonzero correction to scattering requires a derived semiclassical backreaction and an explicit rotating or charged solution.
\end{abstract}

\noindent\textbf{Keywords:} $f(\Qsc)$ gravity; symmetric teleparallel gravity; black-hole thermodynamics; logarithmic entropy correction; Hawking temperature; heat capacity; superradiance

\tableofcontents

\section{Introduction}\label{sec:introduction}
Black-hole thermodynamics connects a geometric horizon with temperature, entropy, and conservation laws. For a Schwarzschild black hole, the standard relations are
\begin{equation}\label{eq:intro-classical}
  S_0=\frac{A_H}{4G},
  \qquad
  T_H=\frac{\kappa_H}{2\pi},
\end{equation}
where $A_H$ is the horizon area and $\kappa_H$ is the surface gravity \cite{Bekenstein1973,Hawking1975}. Many quantum or statistical calculations predict a subleading logarithmic term at large entropy \cite{Carlip2000,Das2002}. A convenient phenomenological form is
\begin{equation}\label{eq:intro-log}
  S_{\mathrm{corr}}
  =S_0+\alpha\ln\!\left(\frac{S_0}{S_\star}\right)
  +\Ord(\alpha^2),
\end{equation}
where $S_\star>0$ makes the logarithm dimensionless.

Three logically different operations must not be conflated:
\begin{enumerate}[label=(\roman*)]
  \item changing the entropy functional;
  \item changing the thermodynamic variable conjugate to that entropy;
  \item changing the spacetime geometry and hence the Hawking radiation problem.
\end{enumerate}
A correction of type (i) does not, by itself, determine corrections of type (ii) or (iii). In particular, a term $\alpha\ln S_0$ is not a semiclassical field equation and therefore cannot by itself determine $A_1(r)$, $B_1(r)$, a shifted horizon, or a corrected wave potential.

This distinction is especially important in $f(\Qsc)$ gravity. The metric and a flat, torsion-free affine connection are independent variables, and a candidate background must satisfy both the metric equation and the connection equation \cite{Jimenez2020,Dambrosio2022,Wang2022}. An arbitrary Schwarzschild-like deformation cannot be combined consistently with an unrelated function $f(\Qsc)$.

We therefore adopt a deliberately controlled benchmark:
\begin{enumerate}
  \item the covariant symmetric-teleparallel framework is stated completely;
  \item the explicit background is the exact Schwarzschild solution of the linear STEGR branch $f(\Qsc)=\Qsc$;
  \item the leading entropy is the branch-specific area entropy;
  \item the logarithmic term modifies only the entropy unless an additional backreaction equation is supplied;
  \item all thermodynamic and scattering claims are obtained from displayed equations rather than schematic plots.
\end{enumerate}

The derivations are organized as follows. \Cref{sec:geometry} develops the geometry and field equations. \Cref{sec:schwarzschild} solves the vacuum equations and verifies a compatible inertial connection. \Cref{sec:classical-thermo} derives the horizon temperature and leading entropy. \Cref{sec:log-thermo} derives the logarithmically corrected thermodynamics and its consistency conditions. \Cref{sec:stability} derives heat capacities and a free-energy diagnostic. \Cref{sec:numerics} evaluates the formulas reproducibly. \Cref{sec:scattering} derives scalar scattering and the no-superradiance result.

\section{Symmetric teleparallel geometry and the covariant field equations}\label{sec:geometry}

\subsection{Independent metric and affine connection}
We work on a four-dimensional manifold equipped with
\begin{equation}
  g_{\mu\nu},
  \qquad
  \Gamma^\alpha{}_{\mu\nu},
\end{equation}
which are varied independently. The curvature and torsion of the affine connection are
\begin{align}
  R^\alpha{}_{\beta\mu\nu}(\Gamma)
  &\equiv
  \partial_\mu\Gamma^\alpha{}_{\nu\beta}
  -\partial_\nu\Gamma^\alpha{}_{\mu\beta}
  +\Gamma^\alpha{}_{\mu\lambda}\Gamma^\lambda{}_{\nu\beta}
  -\Gamma^\alpha{}_{\nu\lambda}\Gamma^\lambda{}_{\mu\beta},
  \label{eq:affine-curvature}\\
  T^\alpha{}_{\mu\nu}(\Gamma)
  &\equiv 2\Gamma^\alpha{}_{[\mu\nu]}.
  \label{eq:torsion}
\end{align}
Symmetric teleparallel geometry imposes
\begin{equation}\label{eq:STG-constraints}
  R^\alpha{}_{\beta\mu\nu}(\Gamma)=0,
  \qquad
  T^\alpha{}_{\mu\nu}(\Gamma)=0.
\end{equation}
The affine connection is therefore flat and symmetric, but it is not required to be metric-compatible.

The nonmetricity tensor is
\begin{align}
  Q_{\alpha\mu\nu}
  &\equiv \nabla_\alpha g_{\mu\nu}
  \nonumber\\
  &=\partial_\alpha g_{\mu\nu}
  -\Gamma^\lambda{}_{\alpha\mu}g_{\lambda\nu}
  -\Gamma^\lambda{}_{\alpha\nu}g_{\mu\lambda}.
  \label{eq:Q-definition-expanded}
\end{align}
Because $g_{\mu\nu}=g_{\nu\mu}$, one has
\begin{equation}
  Q_{\alpha\mu\nu}=Q_{\alpha\nu\mu}.
\end{equation}
The two independent contractions are
\begin{equation}\label{eq:Q-traces}
  Q_\alpha\equiv Q_{\alpha\mu}{}^{\mu}
  =g^{\mu\nu}Q_{\alpha\mu\nu},
  \qquad
  \widetilde Q_\alpha\equiv Q^{\mu}{}_{\alpha\mu}
  =g^{\mu\nu}Q_{\nu\alpha\mu}.
\end{equation}

\subsection{Disformation and decomposition of the connection}
The Levi--Civita connection of the metric is
\begin{equation}\label{eq:LC-connection}
  \mathring{\Gamma}^\alpha{}_{\mu\nu}
  =\frac12g^{\alpha\beta}
  \left(
  \partial_\mu g_{\beta\nu}
  +\partial_\nu g_{\beta\mu}
  -\partial_\beta g_{\mu\nu}
  \right).
\end{equation}
For vanishing torsion, the difference between the affine and Levi--Civita connections is the disformation. With the convention used here,
\begin{equation}\label{eq:disformation}
  L^\alpha{}_{\mu\nu}
  \equiv
  \frac12Q^\alpha{}_{\mu\nu}
  -Q_{(\mu\nu)}{}^\alpha,
\end{equation}
and
\begin{equation}\label{eq:connection-decomposition}
  \Gamma^\alpha{}_{\mu\nu}
  =\mathring{\Gamma}^\alpha{}_{\mu\nu}+L^\alpha{}_{\mu\nu}.
\end{equation}
To verify \eqref{eq:connection-decomposition}, insert it into \eqref{eq:Q-definition-expanded}. Since $\LCnabla_\alpha g_{\mu\nu}=0$,
\begin{align}
  Q_{\alpha\mu\nu}
  &=\nabla_\alpha g_{\mu\nu}
  \nonumber\\
  &=\LCnabla_\alpha g_{\mu\nu}
  -L^\lambda{}_{\alpha\mu}g_{\lambda\nu}
  -L^\lambda{}_{\alpha\nu}g_{\mu\lambda}
  \nonumber\\
  &=-L_{\nu\alpha\mu}-L_{\mu\alpha\nu}.
  \label{eq:Q-from-L}
\end{align}
Solving this algebraic equation for $L^\alpha{}_{\mu\nu}$ gives \eqref{eq:disformation}.

\subsection{Superpotential and explicit quadratic form of \texorpdfstring{$\Qsc$}{Q}}
The nonmetricity conjugate, or superpotential, is defined by \cite{Jimenez2020}
\begin{equation}\label{eq:P-definition}
  P^\alpha{}_{\mu\nu}
  \equiv
  -\frac12L^\alpha{}_{\mu\nu}
  +\frac14\left(Q^\alpha-\widetilde Q^\alpha\right)g_{\mu\nu}
  -\frac14\delta^\alpha{}_{(\mu}Q_{\nu)}.
\end{equation}
The nonmetricity scalar is
\begin{equation}\label{eq:Qscalar-definition}
  \Qsc\equiv-Q_{\alpha\mu\nu}P^{\alpha\mu\nu}.
\end{equation}
Substituting \eqref{eq:P-definition} into \eqref{eq:Qscalar-definition}, inserting the disformation \eqref{eq:disformation}, using the symmetry
$Q_{\alpha\mu\nu}=Q_{\alpha\nu\mu}$, and collecting the four independent quadratic contractions gives
\begin{align}
  \Qsc
  =&\frac14Q_{\alpha\mu\nu}Q^{\alpha\mu\nu}
  -\frac12Q_{\alpha\mu\nu}Q^{\mu\alpha\nu}
  -\frac14Q_\alpha Q^\alpha
  +\frac12Q_\alpha\widetilde Q^\alpha.
  \label{eq:Qscalar-expanded}
\end{align}
A useful way to check every coefficient is to rewrite the same result as
\begin{equation}
  \Qsc
  =-\frac14\left(
  -Q_{\alpha\mu\nu}Q^{\alpha\mu\nu}
  +2Q_{\alpha\mu\nu}Q^{\mu\alpha\nu}
  +Q_\alpha Q^\alpha
  -2Q_\alpha\widetilde Q^\alpha
  \right).
\end{equation}
This expression is consistent with the defining contraction
$\Qsc=-Q_{\alpha\mu\nu}P^{\alpha\mu\nu}$ and with the sign of the gravitational action used below. Equivalently, $P$ is the algebraic conjugate of $Q$:
\begin{equation}\label{eq:P-conjugate}
  P^{\alpha\mu\nu}
  =-\frac12\frac{\partial\Qsc}{\partial Q_{\alpha\mu\nu}}.
\end{equation}

\subsection{Action and definitions of \texorpdfstring{$f_{\Qsc}$}{fQ} and \texorpdfstring{$f_{\Qsc\Qsc}$}{fQQ}}
The action is
\begin{equation}\label{eq:fQ-action}
  I[g,\Gamma,\Psi]
  =\int\dd^4x\sqrt{-g}
  \left[-\frac{1}{16\pi G}f(\Qsc)+\Lag_m(g,\Gamma,\Psi)\right].
\end{equation}
We define
\begin{equation}\label{eq:f-derivatives}
  f_{\Qsc}\equiv\frac{\dd f}{\dd\Qsc},
  \qquad
  f_{\Qsc\Qsc}\equiv\frac{\dd^2f}{\dd\Qsc^2}.
\end{equation}
By the chain rule,
\begin{equation}\label{eq:chain-rule-fQ}
  \nabla_\alpha f_{\Qsc}
  =f_{\Qsc\Qsc}\nabla_\alpha\Qsc.
\end{equation}
Thus $f_{\Qsc\Qsc}$ controls the genuinely nonlinear response to gradients of $\Qsc$.

\subsection{Metric variation: displayed intermediate steps}
At fixed affine connection, the gravitational variation is\cite{Bahamonde2022}
\begin{align}
  \delta I_g
  &=-\frac{1}{16\pi G}\int\dd^4x
  \left[
  \delta(\sqrt{-g})f
  +\sqrt{-g}f_{\Qsc}\delta\Qsc
  \right].
  \label{eq:metric-variation-start}
\end{align}
The determinant varies as
\begin{equation}\label{eq:det-variation}
  \delta\sqrt{-g}
  =-\frac12\sqrt{-g}\,g_{\mu\nu}\delta g^{\mu\nu}.
\end{equation}
The variation of $\Qsc$ contains a derivative of $\delta g^{\mu\nu}$ and algebraic terms. After substituting \eqref{eq:P-conjugate}, using
\begin{equation}
  \delta Q_{\alpha\mu\nu}
  =\nabla_\alpha\delta g_{\mu\nu}
  \qquad(\delta\Gamma^\alpha{}_{\mu\nu}=0),
\end{equation}
and converting $\delta g_{\mu\nu}=-g_{\mu\rho}g_{\nu\sigma}\delta g^{\rho\sigma}$, one obtains
\begin{align}
  \sqrt{-g}f_{\Qsc}\delta\Qsc
  =&-2\sqrt{-g}f_{\Qsc}P^\alpha{}_{\mu\nu}
  \nabla_\alpha\delta g^{\mu\nu}
  \nonumber\\
  &+\sqrt{-g}f_{\Qsc}
  \left(
  P_{\mu\alpha\beta}Q_\nu{}^{\alpha\beta}
  -2Q_{\alpha\beta\mu}P^{\alpha\beta}{}_{\nu}
  \right)\delta g^{\mu\nu}.
  \label{eq:deltaQ-identity}
\end{align}
Integrating the first term by parts gives
\begin{align}
  &-2\int\dd^4x\sqrt{-g}f_{\Qsc}P^\alpha{}_{\mu\nu}
  \nabla_\alpha\delta g^{\mu\nu}
  \nonumber\\
  =&\;2\int\dd^4x\,
  \nabla_\alpha\!\left(\sqrt{-g}f_{\Qsc}P^\alpha{}_{\mu\nu}\right)
  \delta g^{\mu\nu}
  +\text{boundary term}.
  \label{eq:metric-IBP}
\end{align}
The matter stress tensor is defined by
\begin{equation}\label{eq:stress-definition}
  T_{\mu\nu}
  \equiv-\frac{2}{\sqrt{-g}}
  \frac{\delta(\sqrt{-g}\Lag_m)}{\delta g^{\mu\nu}},
\end{equation}
so that
\begin{equation}
  \delta I_m
  =-\frac12\int\dd^4x\sqrt{-g}\,
  T_{\mu\nu}\delta g^{\mu\nu}.
\end{equation}
Combining \eqref{eq:metric-variation-start}--\eqref{eq:metric-IBP}, discarding the boundary variation, and requiring $\delta I=0$ for arbitrary $\delta g^{\mu\nu}$ yields
\begin{align}
  \frac{2}{\sqrt{-g}}\nabla_\alpha
  \left(\sqrt{-g}f_{\Qsc}P^\alpha{}_{\mu\nu}\right)
  +\frac12g_{\mu\nu}f
  +f_{\Qsc}\left(
  P_{\mu\alpha\beta}Q_\nu{}^{\alpha\beta}
  -2Q_{\alpha\beta\mu}P^{\alpha\beta}{}_{\nu}
  \right)
  =8\pi G T_{\mu\nu}.
  \label{eq:metric-field-equation-detailed}
\end{align}

\subsection{Connection variation}
A flat and torsion-free connection is locally pure gauge. It can be written in terms of four functions $\xi^\alpha(x)$ as
\begin{equation}\label{eq:pure-gauge}
  \Gamma^\alpha{}_{\mu\nu}
  =\frac{\partial x^\alpha}{\partial\xi^\rho}
  \partial_\mu\partial_\nu\xi^\rho.
\end{equation}
An allowed infinitesimal variation preserves flatness and torsionlessness and takes the form
\begin{equation}\label{eq:connection-variation}
  \delta\Gamma^\alpha{}_{\mu\nu}
  =\nabla_\mu\nabla_\nu\zeta^\alpha.
\end{equation}
After an irrelevant sign redefinition of the gauge parameter $\zeta^\alpha$.
At fixed metric,
\begin{align}
  \delta Q_{\alpha\mu\nu}
  &=-\delta\Gamma^\lambda{}_{\alpha\mu}g_{\lambda\nu}
  -\delta\Gamma^\lambda{}_{\alpha\nu}g_{\mu\lambda}
  \nonumber\\
  &=-2g_{\lambda(\nu}\delta\Gamma^\lambda{}_{\mu)\alpha}.
  \label{eq:deltaQ-connection}
\end{align}
Using \eqref{eq:P-conjugate}, the connection part of the variation is proportional to
\begin{align}
  \delta_\Gamma I_g
  &\propto
  \int\dd^4x\sqrt{-g}\,
  f_{\Qsc}P^{\mu\nu}{}_{\alpha}
  \nabla_\mu\nabla_\nu\zeta^\alpha.
\end{align}
Integrating by parts twice gives
\begin{align}
  \delta_\Gamma I_g
  &\propto
  \int\dd^4x\,
  \zeta^\alpha
  \nabla_\mu\nabla_\nu
  \left(\sqrt{-g}f_{\Qsc}P^{\mu\nu}{}_{\alpha}\right)
  +\text{boundary terms}.
\end{align}
In the absence of matter hypermomentum, arbitrary $\zeta^\alpha$ therefore gives
\begin{equation}\label{eq:connection-field-equation-detailed}
  \nabla_\mu\nabla_\nu
  \left(\sqrt{-g}f_{\Qsc}P^{\mu\nu}{}_{\alpha}\right)=0.
\end{equation}
Equations \eqref{eq:metric-field-equation-detailed} and \eqref{eq:connection-field-equation-detailed} must be solved together in a nonlinear $f(\Qsc)$ theory.

\subsection{Einstein-tensor form and the STEGR limit}
The metric equation can be rearranged into a form involving the Levi--Civita Einstein tensor:
\begin{equation}\label{eq:Einstein-form-fQ}
  f_{\Qsc}\LCG_{\mu\nu}
  +\frac12g_{\mu\nu}\left(\Qsc f_{\Qsc}-f\right)
  +2f_{\Qsc\Qsc}P^\alpha{}_{\mu\nu}\LCnabla_\alpha\Qsc
  =8\pi G T_{\mu\nu}.
\end{equation}
This equation makes the roles of $f_{\Qsc}$ and $f_{\Qsc\Qsc}$ transparent. For
\begin{equation}\label{eq:STEGR-choice}
  f(\Qsc)=\Qsc,
  \qquad
  f_{\Qsc}=1,
  \qquad
  f_{\Qsc\Qsc}=0,
\end{equation}
we have
\begin{equation}
  \Qsc f_{\Qsc}-f=\Qsc-\Qsc=0,
\end{equation}
and \eqref{eq:Einstein-form-fQ} becomes
\begin{equation}\label{eq:Einstein-equation}
  \LCG_{\mu\nu}=8\pi G T_{\mu\nu}.
\end{equation}

The same conclusion follows directly from the curvature decomposition. Substituting \eqref{eq:connection-decomposition} into the affine Ricci scalar gives
\begin{equation}\label{eq:curvature-identity}
  R(\Gamma)
  =\LCR(g)+\Qsc
  +\LCnabla_\alpha\left(Q^\alpha-\widetilde Q^\alpha\right).
\end{equation}
Since $R(\Gamma)=0$ in symmetric teleparallel geometry,
\begin{equation}\label{eq:R-Q-boundary}
  -\Qsc
  =\LCR(g)
  +\LCnabla_\alpha\left(Q^\alpha-\widetilde Q^\alpha\right).
\end{equation}
Therefore the STEGR gravitational action is
\begin{align}
  I_{\mathrm{STEGR}}
  &=-\frac{1}{16\pi G}\int\dd^4x\sqrt{-g}\,\Qsc
  \nonumber\\
  &=\frac{1}{16\pi G}\int\dd^4x\sqrt{-g}\,\LCR
  +\frac{1}{16\pi G}\int\dd^4x\sqrt{-g}\,
  \LCnabla_\alpha(Q^\alpha-\widetilde Q^\alpha).
  \label{eq:STEGR-EH}
\end{align}
The second integral is a boundary term. Hence the bulk metric equation is exactly Einstein's equation, while the independent inertial connection does not introduce a new propagating degree of freedom on this linear branch.

\section{Explicit Schwarzschild solution and compatible connection}\label{sec:schwarzschild}

\subsection{Solving the vacuum metric equations step by step}
We solve \eqref{eq:Einstein-equation} in vacuum,
\begin{equation}
  T_{\mu\nu}=0,
  \qquad
  \LCG_{\mu\nu}=0.
\end{equation}
Use the general static, spherically symmetric areal-radius ansatz
\begin{equation}\label{eq:general-SSS}
  \dd s^2
  =-e^{2\delta(r)}N(r)\dd t^2
  +\frac{\dd r^2}{N(r)}
  +r^2\dd\Omega_2^2,
  \qquad
  \dd\Omega_2^2=\dd\theta^2+\sin^2\theta\,\dd\phi^2.
\end{equation}
A direct calculation gives the mixed Einstein-tensor components
\begin{align}
  \LCG^t{}_t
  &=\frac{rN'+N-1}{r^2},
  \label{eq:Gtt}\\
  \LCG^r{}_r
  &=\frac{rN'+N-1}{r^2}+\frac{2N\delta'}{r},
  \label{eq:Grr}\\
  \LCG^\theta{}_\theta
  &=N(\delta')^2+N\delta''
  +\frac32N'\delta'
  +\frac12N''
  +\frac{N\delta'}{r}
  +\frac{N'}{r},
  \label{eq:Gtheta}
\end{align}
with $\LCG^\phi{}_\phi=\LCG^\theta{}_\theta$.

The $t$--$t$ equation is
\begin{equation}
  rN'+N-1=0.
\end{equation}
Since
\begin{equation}
  \frac{\dd}{\dd r}(rN)=N+rN',
\end{equation}
this equation is
\begin{equation}
  \frac{\dd}{\dd r}(rN)=1.
\end{equation}
Integrating once,
\begin{equation}
  rN=r+C_1,
\end{equation}
so
\begin{equation}\label{eq:N-general}
  N(r)=1+\frac{C_1}{r}.
\end{equation}

Subtracting \eqref{eq:Gtt} from \eqref{eq:Grr} gives
\begin{equation}
  \LCG^r{}_r-\LCG^t{}_t
  =\frac{2N\delta'}{r}=0.
\end{equation}
Outside a horizon $N\neq0$, hence
\begin{equation}
  \delta'(r)=0,
  \qquad
  \delta(r)=\delta_0.
\end{equation}
The constant $e^{2\delta_0}$ is removed by the time rescaling
\begin{equation}
  t\longrightarrow e^{\delta_0}t.
\end{equation}
Thus we set $\delta_0=0$.

Asymptotic flatness requires $N\to1$ as $r\to\infty$, which is already satisfied. Matching the Newtonian potential
\begin{equation}
  g_{tt}\simeq-(1+2\Phi_N),
  \qquad
  \Phi_N=-\frac{GM}{r},
\end{equation}
with $g_{tt}=-N$ gives
\begin{equation}
  C_1=-2GM.
\end{equation}
Therefore
\begin{equation}\label{eq:Schwarzschild-F}
  F(r)\equiv N(r)=1-\frac{2GM}{r},
\end{equation}
and
\begin{equation}\label{eq:Schwarzschild-line-element}
  \dd s^2
  =-F(r)\dd t^2+\frac{\dd r^2}{F(r)}+r^2\dd\Omega_2^2.
\end{equation}
Finally, substituting $\delta'=0$ and $N=1+C_1/r$ into \eqref{eq:Gtheta} gives
\begin{equation}
  \LCG^\theta{}_\theta
  =\frac12N''+\frac{N'}{r}
  =\frac12\frac{2C_1}{r^3}-\frac{C_1}{r^3}=0,
\end{equation}
so the angular equation is satisfied.

\subsection{Horizon and mass--radius relation}
The event horizon is the largest positive root of $F(r)=0$:
\begin{align}
  1-\frac{2GM}{r_+}&=0,
  \\
  r_+&=2GM.
  \label{eq:rplus}
\end{align}
Solving for the mass gives
\begin{equation}\label{eq:M-rplus}
  M(r_+)=\frac{r_+}{2G},
  \qquad
  \frac{\dd M}{\dd r_+}=\frac{1}{2G}.
\end{equation}
The horizon area is
\begin{equation}\label{eq:horizon-area}
  A_H=\int_0^{2\pi}\dd\phi\int_0^\pi\dd\theta\,
  r_+^2\sin\theta
  =4\pi r_+^2.
\end{equation}

\subsection{Why coincident gauge is not imposed naively in the spherical chart}
The coincident gauge is defined by coordinates $\xi^\alpha$ in which
\begin{equation}
  \Gamma^\alpha{}_{\mu\nu}=0.
\end{equation}
After a nonlinear coordinate transformation, the connection is generally nonzero. A convenient affine-coordinate map from spherical coordinates is
\begin{equation}\label{eq:affine-map}
  \xi^0=t,
  \quad
  \xi^1=r\sin\theta\cos\phi,
  \quad
  \xi^2=r\sin\theta\sin\phi,
  \quad
  \xi^3=r\cos\theta.
\end{equation}
Substitution into \eqref{eq:pure-gauge} gives the familiar nonzero inertial components
\begin{align}
  \Gamma^r{}_{\theta\theta}&=-r,
  &\Gamma^r{}_{\phi\phi}&=-r\sin^2\theta,
  \\
  \Gamma^\theta{}_{r\theta}
  =\Gamma^\theta{}_{\theta r}&=\frac1r,
  &\Gamma^\theta{}_{\phi\phi}&=-\sin\theta\cos\theta,
  \\
  \Gamma^\phi{}_{r\phi}
  =\Gamma^\phi{}_{\phi r}&=\frac1r,
  &\Gamma^\phi{}_{\theta\phi}
  =\Gamma^\phi{}_{\phi\theta}&=\cot\theta.
  \label{eq:inertial-connection-components}
\end{align}
This connection is flat and torsion-free because it is generated from a vanishing connection by a coordinate transformation. It is therefore a valid representative of the inertial connection in the standard spherical chart.

\subsection{Explicit nonmetricity check for the benchmark pair}
For the metric \eqref{eq:Schwarzschild-line-element} and the inertial connection \eqref{eq:inertial-connection-components}, the nonzero components of $Q_{\alpha\mu\nu}$ are
\begin{align}
  Q_{rtt}&=-F',
  &Q_{rrr}&=-\frac{F'}{F^2},
  \\
  Q_{\theta r\theta}=Q_{\theta\theta r}
  &=\frac{r(1-F)}{F},
  &Q_{\phi r\phi}=Q_{\phi\phi r}
  &=\frac{r(1-F)\sin^2\theta}{F}.
  \label{eq:explicit-Q-components}
\end{align}
The first trace vanishes:
\begin{equation}
  Q_\alpha=0.
\end{equation}
The second trace has one nonzero component,
\begin{equation}\label{eq:Qtilde-r}
  \widetilde Q_r
  =\frac{2-2F-rF'}{rF}.
\end{equation}
The two quadratic contractions are
\begin{align}
  Q_{\alpha\mu\nu}Q^{\alpha\mu\nu}
  &=\frac{2\left[r^2(F')^2+2(F-1)^2\right]}{r^2F},
  \\
  Q_{\alpha\mu\nu}Q^{\mu\alpha\nu}
  &=\frac{r^2(F')^2+2(F-1)^2}{r^2F}.
\end{align}
Using \eqref{eq:Qscalar-expanded} and $Q_\alpha=0$,
\begin{align}
  \Qsc
  &=\frac14
  \frac{2\left[r^2(F')^2+2(F-1)^2\right]}{r^2F}
  -\frac12
  \frac{r^2(F')^2+2(F-1)^2}{r^2F}
  \\
  &=0.
  \label{eq:Qzero-branch}
\end{align}
This explicit calculation illustrates why the metric and connection must be specified together. It also provides a direct constant-$\Qsc$ representative for the Schwarzschild branch. The vacuum metric equation is nevertheless obtained from the full covariant variation, not by prematurely inserting a symmetry-reduced ansatz into the action.

For the linear STEGR theory, the connection equation is redundant because the connection dependence in \eqref{eq:STEGR-EH} is confined to a boundary term. Thus the pair \eqref{eq:Schwarzschild-line-element} and \eqref{eq:inertial-connection-components} satisfies both sectors of the theory.

\section{Classical horizon temperature and leading entropy}\label{sec:classical-thermo}

\subsection{Surface gravity from the Killing-vector definition}
The static Killing vector is
\begin{equation}
  \chi^\mu=(1,0,0,0),
  \qquad
  \chi_\mu=g_{\mu\nu}\chi^\nu=(-F,0,0,0).
\end{equation}
For the metric \eqref{eq:Schwarzschild-line-element}, the relevant Levi--Civita symbols are
\begin{equation}
  \mathring\Gamma^t{}_{tr}
  =\mathring\Gamma^t{}_{rt}=\frac{F'}{2F},
  \qquad
  \mathring\Gamma^r{}_{tt}=\frac12FF',
  \qquad
  \mathring\Gamma^r{}_{rr}=-\frac{F'}{2F}.
\end{equation}
Now
\begin{align}
  \LCnabla_r\chi_t
  &=\partial_r(-F)-\mathring\Gamma^t{}_{rt}\chi_t
  \\
  &=-F'-\frac{F'}{2F}(-F)
  \\
  &=-\frac12F',
  \label{eq:nabla-r-chi-t}
\end{align}
and
\begin{align}
  \LCnabla_t\chi_r
  &=-\mathring\Gamma^t{}_{tr}\chi_t
  \\
  &=-\frac{F'}{2F}(-F)
  \\
  &=\frac12F'.
  \label{eq:nabla-t-chi-r}
\end{align}
The surface gravity is defined by
\begin{equation}\label{eq:kappa-definition}
  \kappa_H^2
  =-\frac12
  (\LCnabla_\mu\chi_\nu)(\LCnabla^\mu\chi^\nu)
  \bigg|_{r=r_+}.
\end{equation}
The two nonzero products are equal:
\begin{align}
  (\LCnabla_r\chi_t)(\LCnabla^r\chi^t)
  &=-\frac{(F')^2}{4},
  \\
  (\LCnabla_t\chi_r)(\LCnabla^t\chi^r)
  &=-\frac{(F')^2}{4}.
\end{align}
Hence
\begin{equation}
  (\LCnabla_\mu\chi_\nu)(\LCnabla^\mu\chi^\nu)
  =-\frac{(F')^2}{2},
\end{equation}
and
\begin{equation}\label{eq:kappa-Fprime}
  \kappa_H=\frac12F'(r_+).
\end{equation}
Since
\begin{equation}
  F'(r)=\frac{2GM}{r^2},
\end{equation}
and $2GM=r_+$,
\begin{equation}
  F'(r_+)=\frac{r_+}{r_+^2}=\frac1{r_+}.
\end{equation}
Therefore
\begin{equation}\label{eq:kappa-final}
  \kappa_H=\frac1{2r_+}.
\end{equation}
The geometric Hawking temperature is
\begin{equation}\label{eq:T-kappa-final}
  \Tsg=\frac{\kappa_H}{2\pi}
  =\frac1{4\pi r_+}.
\end{equation}

\subsection{Independent Euclidean regularity check}
Set $t=-i\tau$. Near the horizon,
\begin{equation}
  F(r)=F'(r_+)(r-r_+)+\Ord((r-r_+)^2).
\end{equation}
The $(\tau,r)$ part of the Euclidean metric is
\begin{equation}
  \dd s_E^2
  \simeq F'(r_+)(r-r_+)\dd\tau^2
  +\frac{\dd r^2}{F'(r_+)(r-r_+)}.
\end{equation}
Define
\begin{equation}
  \rho\equiv2\sqrt{\frac{r-r_+}{F'(r_+)}}.
\end{equation}
Then
\begin{equation}
  r-r_+=\frac14F'(r_+)\rho^2,
  \qquad
  \dd r=\frac12F'(r_+)\rho\,\dd\rho.
\end{equation}
Substitution gives
\begin{equation}
  \dd s_E^2
  \simeq\dd\rho^2
  +\left(\frac{F'(r_+)}2\right)^2\rho^2\dd\tau^2.
\end{equation}
This is a regular plane in polar coordinates only if
\begin{equation}
  \frac{F'(r_+)}2\tau
  \sim
  \frac{F'(r_+)}2\tau+2\pi.
\end{equation}
Thus the Euclidean period is
\begin{equation}
  \beta_H=\frac{4\pi}{F'(r_+)},
\end{equation}
and
\begin{equation}
  T_H=\beta_H^{-1}=\frac{F'(r_+)}{4\pi}
  =\frac1{4\pi r_+},
\end{equation}
confirming \eqref{eq:T-kappa-final}.

\subsection{Leading entropy from the first law}
On the STEGR branch, the extended Wald construction reproduces the GR entropy \cite{HeisenbergEntropy2022}. The same expression follows from the classical first law. Using
\begin{equation}
  \dd M=\Tsg\,\dd S_0,
\end{equation}
we obtain
\begin{equation}
  \frac{\dd S_0}{\dd r_+}
  =\frac{1}{\Tsg}\frac{\dd M}{\dd r_+}.
\end{equation}
Substituting \eqref{eq:M-rplus} and \eqref{eq:T-kappa-final},
\begin{align}
  \frac{\dd S_0}{\dd r_+}
  &=\left(4\pi r_+\right)\left(\frac1{2G}\right)
  \\
  &=\frac{2\pi r_+}{G}.
\end{align}
Integrating,
\begin{equation}
  S_0(r_+)
  =\int\frac{2\pi r_+}{G}\dd r_+
  =\frac{\pi r_+^2}{G}+S_{\mathrm{int}}.
\end{equation}
Choosing the additive constant $S_{\mathrm{int}}=0$ gives
\begin{equation}\label{eq:S0-final}
  S_0=\frac{\pi r_+^2}{G}
  =\frac{4\pi r_+^2}{4G}
  =\frac{A_H}{4G}.
\end{equation}
This equality is justified for the benchmark branch. It is not assumed as a universal formula for every nonlinear $f(\Qsc)$ solution.

\section{Logarithmic entropy correction and temperature definitions}\label{sec:log-thermo}

\subsection{Corrected entropy and its derivatives}
We correct the branch-specific leading entropy \eqref{eq:S0-final}:
\begin{equation}\label{eq:S-corr-def}
  \Scorr
  =S_0+\alpha\ln\!\left(\frac{S_0}{S_\star}\right)
  +\Ord(\alpha^2).
\end{equation}
Substituting $S_0=\pi r_+^2/G$ gives
\begin{equation}\label{eq:S-corr-explicit}
  \Scorr(r_+)
  =\frac{\pi r_+^2}{G}
  +\alpha\ln\!\left(\frac{\pi r_+^2}{G S_\star}\right).
\end{equation}
Differentiate the first term:
\begin{equation}
  \frac{\dd}{\dd r_+}\left(\frac{\pi r_+^2}{G}\right)
  =\frac{2\pi r_+}{G}.
\end{equation}
For the logarithm, use $\dd(\ln X)/\dd r=X'/X$:
\begin{align}
  \frac{\dd}{\dd r_+}
  \ln\!\left(\frac{\pi r_+^2}{GS_\star}\right)
  &=\frac{
  \frac{\dd}{\dd r_+}(\pi r_+^2/(GS_\star))
  }{\pi r_+^2/(GS_\star)}
  \\
  &=\frac{2\pi r_+/(GS_\star)}{\pi r_+^2/(GS_\star)}
  \\
  &=\frac2{r_+}.
\end{align}
Therefore
\begin{equation}\label{eq:Sprime}
  \frac{\dd\Scorr}{\dd r_+}
  =\frac{2\pi r_+}{G}+\frac{2\alpha}{r_+}.
\end{equation}
The second derivative is
\begin{equation}\label{eq:Ssecond}
  \frac{\dd^2\Scorr}{\dd r_+^2}
  =\frac{2\pi}{G}-\frac{2\alpha}{r_+^2}.
\end{equation}

\subsection{Perturbative control parameters}
The logarithmic term is subleading when
\begin{equation}\label{eq:epsilon-log}
  \varepsilon_{\log}
  \equiv
  \frac{\left|\alpha\ln(S_0/S_\star)\right|}{S_0}
  \ll1.
\end{equation}
Derivative-based thermodynamics also introduces
\begin{equation}\label{eq:x-def}
  x\equiv\frac{\alpha}{S_0}
  =\frac{\alpha G}{\pi r_+^2},
  \qquad
  |x|\ll1.
\end{equation}
The condition $|x|\ll1$ is essential because $\dd\Scorr/\dd r_+$ contains the relative correction
\begin{equation}
  \frac{2\alpha/r_+}{2\pi r_+/G}
  =\frac{\alpha G}{\pi r_+^2}=x.
\end{equation}

\subsection{Geometric Hawking temperature remains unchanged without backreaction}
The surface-gravity temperature is a functional of the metric:
\begin{equation}
  \Tsg[g]=\frac{F'(r_+)}{4\pi}.
\end{equation}
In this work the entropy correction is not accompanied by a corrected effective action. We therefore impose
\begin{equation}\label{eq:no-backreaction-assumption}
  F(r;\alpha)=F_0(r),
  \qquad
  r_+(\alpha)=r_+^{(0)},
  \qquad
  M(\alpha)=M_0.
\end{equation}
Differentiating with respect to $\alpha$ gives
\begin{equation}
  \frac{\partial F}{\partial\alpha}=0,
  \qquad
  \frac{\partial r_+}{\partial\alpha}=0,
  \qquad
  \frac{\partial M}{\partial\alpha}=0.
\end{equation}
Consequently,
\begin{equation}\label{eq:T-kappa-alpha}
  \frac{\partial\Tsg}{\partial\alpha}=0,
  \qquad
  \Tsg=\frac1{4\pi r_+}.
\end{equation}
This is the physical Hawking temperature of the fixed background.

\subsection{Why the corrected entropy is not compatible with the unmodified first law}
If one simultaneously keeps $M(r_+)$, $\Tsg(r_+)$, and replaces $S_0$ by $\Scorr$, then
\begin{align}
  \Tsg\frac{\dd\Scorr}{\dd r_+}
  &=\frac1{4\pi r_+}
  \left(\frac{2\pi r_+}{G}+\frac{2\alpha}{r_+}\right)
  \\
  &=\frac1{2G}+\frac{\alpha}{2\pi r_+^2}.
  \label{eq:first-law-mismatch-left}
\end{align}
But
\begin{equation}
  \frac{\dd M}{\dd r_+}=\frac1{2G}.
\end{equation}
Hence
\begin{equation}\label{eq:first-law-mismatch}
  \frac{\dd M}{\dd r_+}
  -\Tsg\frac{\dd\Scorr}{\dd r_+}
  =-\frac{\alpha}{2\pi r_+^2}.
\end{equation}
The mismatch is nonzero for $\alpha\neq0$. One must therefore choose which quantity is being redefined.

\subsection{First-law effective temperature with fixed classical mass}
Keeping the classical mass function and defining a conjugate effective temperature by
\begin{equation}\label{eq:first-law-effective}
  \dd M=\Tfl\,\dd\Scorr
\end{equation}
gives
\begin{equation}
  \Tfl
  =\frac{\dd M/\dd r_+}{\dd\Scorr/\dd r_+}.
\end{equation}
Insert the derivatives:
\begin{align}
  \Tfl
  &=\frac{1/(2G)}{2\pi r_+/G+2\alpha/r_+}
  \\
  &=\frac{1}{2G}
  \frac{1}{2(\pi r_+/G+\alpha/r_+)}
  \\
  &=\frac{1}{4G}
  \frac{1}{\pi r_+/G+\alpha/r_+}.
\end{align}
Multiply the denominator by $Gr_+$:
\begin{align}
  \Tfl
  &=\frac{1}{4G}
  \frac{Gr_+}{\pi r_+^2+\alpha G}
  \\
  &=\frac{r_+}{4(\pi r_+^2+\alpha G)}.
  \label{eq:TFL-final}
\end{align}
Factor out $\pi r_+^2$:
\begin{align}
  \Tfl
  &=\frac{r_+}{4\pi r_+^2}
  \frac{1}{1+\alpha G/(\pi r_+^2)}
  \\
  &=\Tsg\frac{1}{1+x}.
  \label{eq:TFL-x}
\end{align}
For $|x|<1$,
\begin{equation}
  \frac1{1+x}=1-x+x^2-x^3+\cdots.
\end{equation}
Therefore
\begin{equation}\label{eq:TFL-series}
  \Tfl
  =\Tsg\left[
  1-\frac{\alpha G}{\pi r_+^2}
  +\Ord\!\left(\frac{\alpha^2G^2}{r_+^4}\right)
  \right].
\end{equation}
Thus, in the controlled region,
\begin{equation}
  \alpha>0\;\Longrightarrow\;\Tfl<\Tsg,
  \qquad
  \alpha<0\;\Longrightarrow\;\Tfl>\Tsg.
\end{equation}

\subsection{Alternative thermodynamic reconstruction if \texorpdfstring{$T_\kappa$}{Tkappa} is held fixed}
Instead of redefining the temperature, one could formally reconstruct a mass function $M_\alpha(r_+)$ from
\begin{equation}
  \frac{\dd M_\alpha}{\dd r_+}
  =\Tsg\frac{\dd\Scorr}{\dd r_+}.
\end{equation}
Using \eqref{eq:first-law-mismatch-left},
\begin{equation}
  \frac{\dd M_\alpha}{\dd r_+}
  =\frac1{2G}+\frac{\alpha}{2\pi r_+^2}.
\end{equation}
Integrating term by term,
\begin{align}
  M_\alpha(r_+)
  &=\int\left(\frac1{2G}+\frac{\alpha}{2\pi r_+^2}\right)\dd r_+
  \\
  &=\frac{r_+}{2G}
  -\frac{\alpha}{2\pi r_+}+C_M.
  \label{eq:formal-Malpha}
\end{align}
Equation \eqref{eq:formal-Malpha} is only a thermodynamic reconstruction. It is not identified with the ADM mass because no corrected metric has been derived. The present paper therefore uses the fixed classical mass and labels \eqref{eq:TFL-final} an \emph{effective first-law temperature}, not a backreacted Hawking temperature.

\section{Heat capacities, free-energy diagnostic, and controlled stability}\label{sec:stability}

\subsection{Geometric heat capacity}
Along the Schwarzschild family,
\begin{equation}
  C_\kappa
  \equiv\frac{\dd M}{\dd\Tsg}
  =\frac{\dd M/\dd r_+}{\dd\Tsg/\dd r_+}.
\end{equation}
From \eqref{eq:T-kappa-final},
\begin{equation}
  \frac{\dd\Tsg}{\dd r_+}
  =-\frac1{4\pi r_+^2}.
\end{equation}
Therefore
\begin{align}
  C_\kappa
  &=\frac{1/(2G)}{-1/(4\pi r_+^2)}
  \\
  &=-\frac{2\pi r_+^2}{G}.
  \label{eq:Ckappa-final}
\end{align}
It is negative for every $r_+>0$.

A different quantity can be formed from the corrected entropy and the geometric temperature:
\begin{align}
  \widehat C_\kappa
  &\equiv\Tsg\frac{\dd\Scorr}{\dd\Tsg}
  \\
  &=\Tsg
  \frac{\dd\Scorr/\dd r_+}{\dd\Tsg/\dd r_+}
  \\
  &=\frac1{4\pi r_+}
  \frac{2\pi r_+/G+2\alpha/r_+}{-1/(4\pi r_+^2)}
  \\
  &=-\frac{2\pi r_+^2}{G}-2\alpha.
  \label{eq:Chat-kappa}
\end{align}
Because the corrected entropy and $T_\kappa$ do not satisfy the unmodified first law, $\widehat C_\kappa$ is not equal to $\dd M/\dd T_\kappa$ and should not be confused with \eqref{eq:Ckappa-final}.

\subsection{Effective first-law heat capacity}
Let
\begin{equation}
  D(r_+)\equiv\pi r_+^2+\alpha G.
\end{equation}
Then
\begin{equation}
  \Tfl=\frac{r_+}{4D}.
\end{equation}
Differentiate using the quotient rule:
\begin{align}
  \frac{\dd\Tfl}{\dd r_+}
  &=\frac{(1)(4D)-r_+(4D')}{(4D)^2}
  \\
  &=\frac{D-r_+D'}{4D^2}.
\end{align}
Since
\begin{equation}
  D'=2\pi r_+,
\end{equation}
we obtain
\begin{align}
  D-r_+D'
  &=\pi r_+^2+\alpha G-2\pi r_+^2
  \\
  &=\alpha G-\pi r_+^2.
\end{align}
Hence
\begin{equation}\label{eq:dTFL-dr}
  \frac{\dd\Tfl}{\dd r_+}
  =\frac{\alpha G-\pi r_+^2}
  {4(\pi r_+^2+\alpha G)^2}.
\end{equation}
The heat capacity is
\begin{align}
  C_{\mathrm{FL}}
  &\equiv\frac{\dd M}{\dd\Tfl}
  =\frac{\dd M/\dd r_+}{\dd\Tfl/\dd r_+}
  \\
  &=\frac{1/(2G)}{
  (\alpha G-\pi r_+^2)/[4(\pi r_+^2+\alpha G)^2]
  }
  \\
  &=\frac{2(\pi r_+^2+\alpha G)^2}
  {G(\alpha G-\pi r_+^2)}.
  \label{eq:CFL-final}
\end{align}
At $\alpha=0$,
\begin{equation}
  C_{\mathrm{FL}}\big|_{\alpha=0}
  =\frac{2\pi^2r_+^4}{G(-\pi r_+^2)}
  =-\frac{2\pi r_+^2}{G}
  =C_\kappa.
\end{equation}

Using $x=\alpha G/(\pi r_+^2)$,
\begin{align}
  C_{\mathrm{FL}}
  &=\frac{2\pi^2r_+^4(1+x)^2}
  {G\pi r_+^2(x-1)}
  \\
  &=-\frac{2\pi r_+^2}{G}
  \frac{(1+x)^2}{1-x}.
  \label{eq:CFL-x}
\end{align}
Expanding,
\begin{equation}
  \frac{(1+x)^2}{1-x}
  =(1+2x+x^2)(1+x+x^2+\cdots)
  =1+3x+\Ord(x^2),
\end{equation}
so
\begin{equation}\label{eq:CFL-series}
  C_{\mathrm{FL}}
  =-\frac{2\pi r_+^2}{G}
  \left[1+3\frac{\alpha G}{\pi r_+^2}
  +\Ord\!\left(\frac{\alpha^2G^2}{r_+^4}\right)
  \right].
\end{equation}

\subsection{Formal pole and breakdown of the expansion}
For $\alpha>0$, the denominator of \eqref{eq:CFL-final} vanishes at
\begin{equation}
  \alpha G-\pi r_c^2=0,
\end{equation}
so
\begin{equation}\label{eq:rc-pole}
  r_c^2=\frac{\alpha G}{\pi}.
\end{equation}
At this radius,
\begin{equation}
  S_0(r_c)=\frac{\pi r_c^2}{G}=\alpha.
\end{equation}
Therefore
\begin{equation}
  \left|\frac{\alpha}{S_0(r_c)}\right|=1.
\end{equation}
But the expansion requires $|\alpha|/S_0\ll1$. The pole occurs exactly outside the controlled domain. It cannot be used as evidence for a second-order phase transition.

In the controlled region $|x|\ll1$, one has $1-x>0$ and $(1+x)^2>0$, so
\begin{equation}
  C_{\mathrm{FL}}<0.
\end{equation}
Thus the logarithmic correction does not produce a reliable stable canonical branch within the perturbative regime.

\subsection{Helmholtz diagnostic}
For the internally consistent pair $(\Scorr,\Tfl)$, define
\begin{equation}\label{eq:F-def}
  \cF_{\mathrm{FL}}
  \equiv M-\Tfl\Scorr.
\end{equation}
Substitution gives
\begin{align}
  \cF_{\mathrm{FL}}(r_+)
  =&\frac{r_+}{2G}
  -\frac{r_+}{4(\pi r_+^2+\alpha G)}
  \left[
  \frac{\pi r_+^2}{G}
  +\alpha\ln\!\left(\frac{\pi r_+^2}{GS_\star}\right)
  \right].
  \label{eq:F-explicit}
\end{align}
At fixed $\alpha$,
\begin{align}
  \dd\cF_{\mathrm{FL}}
  &=\dd M-\Tfl\dd\Scorr-\Scorr\dd\Tfl
  \\
  &=-\Scorr\dd\Tfl,
  \label{eq:F-differential}
\end{align}
where \eqref{eq:first-law-effective} was used. In the controlled region $r_+^2\gg\alpha G/\pi$, $T_{\mathrm{FL}}(r_+)$ is monotonic and the family supplies no pair of competing stable branches at the same temperature. Moreover, asymptotically flat Schwarzschild spacetime has negative heat capacity and does not define the standard stable canonical ensemble without a cavity or AdS boundary. We therefore make no Hawking--Page, reentrant, or triple-point claim.

\section{Reproducible numerical evaluation}\label{sec:numerics}
All values below are direct substitutions into the displayed formulas. Set
\begin{equation}
  G=1,
  \qquad
  S_\star=1,
  \qquad
  r_+=2.
\end{equation}
Then
\begin{equation}
  M=\frac{r_+}{2}=1,
  \qquad
  S_0=\pi r_+^2=4\pi\simeq12.5663706,
\end{equation}
and
\begin{equation}\label{eq:numerical-Tkappa}
  \Tsg=\frac1{4\pi r_+}
  =\frac1{8\pi}
  \simeq0.0397887.
\end{equation}
The value $0.125=1/8$ would omit the factor $\pi$ and is incorrect.

For $\alpha=0.5$,
\begin{align}
  \Scorr
  &=4\pi+0.5\ln(4\pi)
  \\
  &\simeq12.5663706+0.5(2.53102425)
  \\
  &\simeq13.8318827,
\end{align}
\begin{align}
  \Tfl
  &=\frac{2}{4(4\pi+0.5)}
  \\
  &=\frac{1}{2(13.0663706)}
  \\
  &\simeq0.0382662,
\end{align}
and
\begin{align}
  C_{\mathrm{FL}}
  &=\frac{2(4\pi+0.5)^2}{0.5-4\pi}
  \\
  &\simeq-28.2984912.
\end{align}
The other entries are evaluated identically.

\begin{table}[ht]
\centering
\caption{Fully evaluated quantities for $G=S_\star=1$ and $r_+=2$. The geometric quantities $M$, $r_+$, and $T_\kappa$ are independent of $\alpha$ because no backreaction is assumed.}
\label{tab:values}
\begin{tabular}{@{}rrrrrr@{}}
\toprule
$\alpha$ & $M$ & $T_\kappa$ & $S_{\mathrm{corr}}$ & $T_{\mathrm{FL}}$ & $C_{\mathrm{FL}}$\\
\midrule
$-0.5$ & $1.000000$ & $0.039789$ & $11.300858$ & $0.041437$ & $-22.285806$\\
$0$    & $1.000000$ & $0.039789$ & $12.566371$ & $0.039789$ & $-25.132741$\\
$+0.5$ & $1.000000$ & $0.039789$ & $13.831883$ & $0.038266$ & $-28.298491$\\
\bottomrule
\end{tabular}
\end{table}

For $|\alpha|=0.5$,
\begin{align}
  \varepsilon_{\log}
  &=\frac{0.5\ln(4\pi)}{4\pi}
  \\
  &\simeq0.1007,
\end{align}
while
\begin{equation}
  |x|=\frac{0.5}{4\pi}\simeq0.0398.
\end{equation}
The example is suitable for illustrating the sign and scale of the correction, although a more precise asymptotic analysis would use still larger $S_0$.

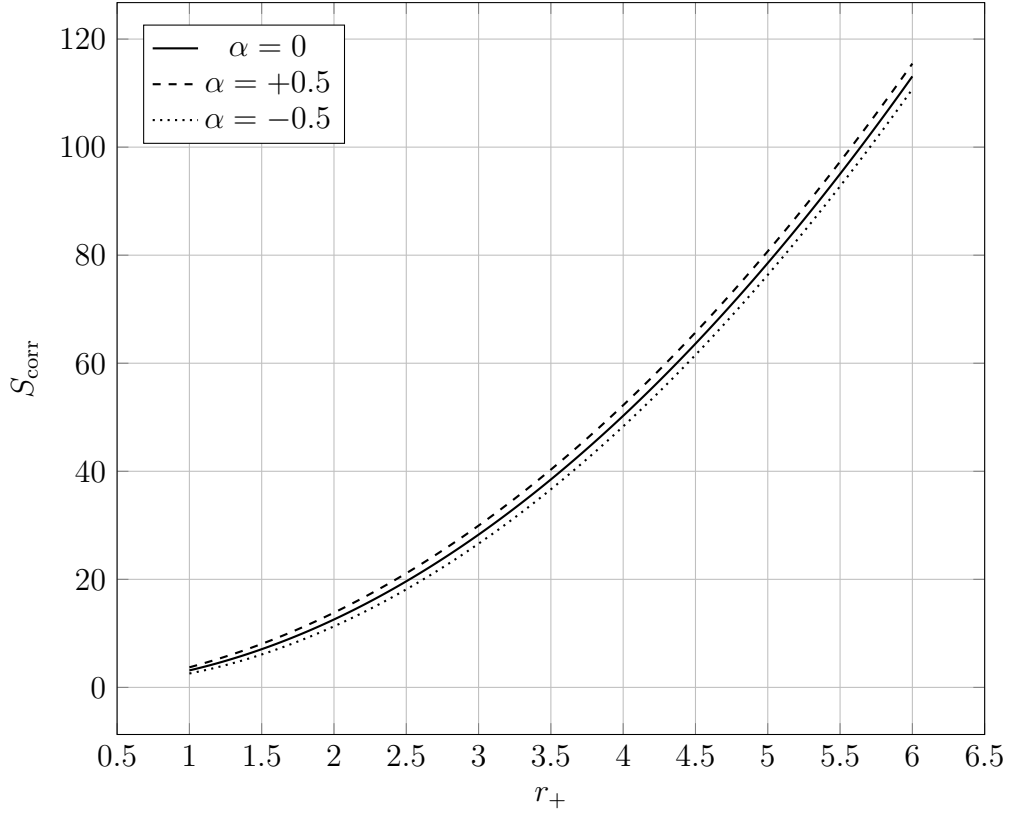
\begin{figure}[ht]
\centering
\begin{tikzpicture}
\begin{axis}[
  width=0.82\textwidth,
  xlabel={$r_+$},
  ylabel={$S_{\mathrm{corr}}$},
  domain=1:6,
  samples=200,
  legend pos=north west,
  grid=major
]
\addplot[thick] {pi*x^2};
\addlegendentry{$\alpha=0$}
\addplot[thick,dashed] {pi*x^2+0.5*ln(pi*x^2)};
\addlegendentry{$\alpha=+0.5$}
\addplot[thick,dotted] {pi*x^2-0.5*ln(pi*x^2)};
\addlegendentry{$\alpha=-0.5$}
\end{axis}
\end{tikzpicture}
\caption{Direct evaluation of \eqref{eq:S-corr-explicit} with $G=S_\star=1$.}
\label{fig:entropy}
\end{figure}

\begin{figure}[ht]
\centering
\begin{tikzpicture}
\begin{axis}[
  width=0.82\textwidth,
  xlabel={$r_+$},
  ylabel={$T$},
  domain=1:8,
  samples=200,
  legend pos=north east,
  grid=major,
  ymin=0
]
\addplot[thick] {1/(4*pi*x)};
\addlegendentry{$T_\kappa$}
\addplot[thick,dashed] {x/(4*(pi*x^2+0.5))};
\addlegendentry{$T_{\mathrm{FL}},\ \alpha=+0.5$}
\addplot[thick,dotted] {x/(4*(pi*x^2-0.5))};
\addlegendentry{$T_{\mathrm{FL}},\ \alpha=-0.5$}
\end{axis}
\end{tikzpicture}
\caption{The geometric temperature and first-law effective temperature. Positive $\alpha$ lowers $T_{\mathrm{FL}}$ in the controlled domain.}
\label{fig:temperature}
\end{figure}
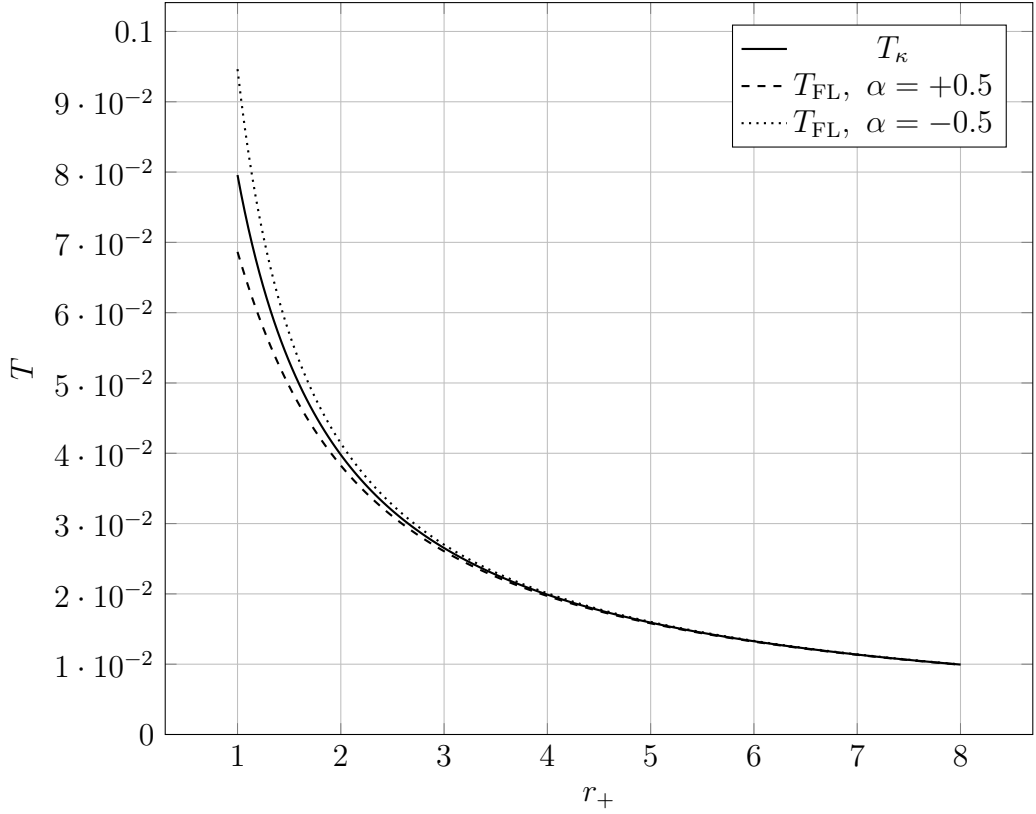

\begin{figure}[ht]
\centering
\begin{tikzpicture}
\begin{axis}[
  width=0.82\textwidth,
  xlabel={$r_+$},
  ylabel={$C_{\mathrm{FL}}$},
  domain=1:6,
  samples=200,
  legend pos=south west,
  grid=major
]
\addplot[thick] {-2*pi*x^2};
\addlegendentry{$\alpha=0$}
\addplot[thick,dashed] {2*(pi*x^2+0.5)^2/(0.5-pi*x^2)};
\addlegendentry{$\alpha=+0.5$}
\addplot[thick,dotted] {2*(pi*x^2-0.5)^2/(-0.5-pi*x^2)};
\addlegendentry{$\alpha=-0.5$}
\end{axis}
\end{tikzpicture}
\caption{The effective heat capacity \eqref{eq:CFL-final} in a perturbative domain. All displayed branches remain negative.}
\label{fig:heat-capacity}
\end{figure}
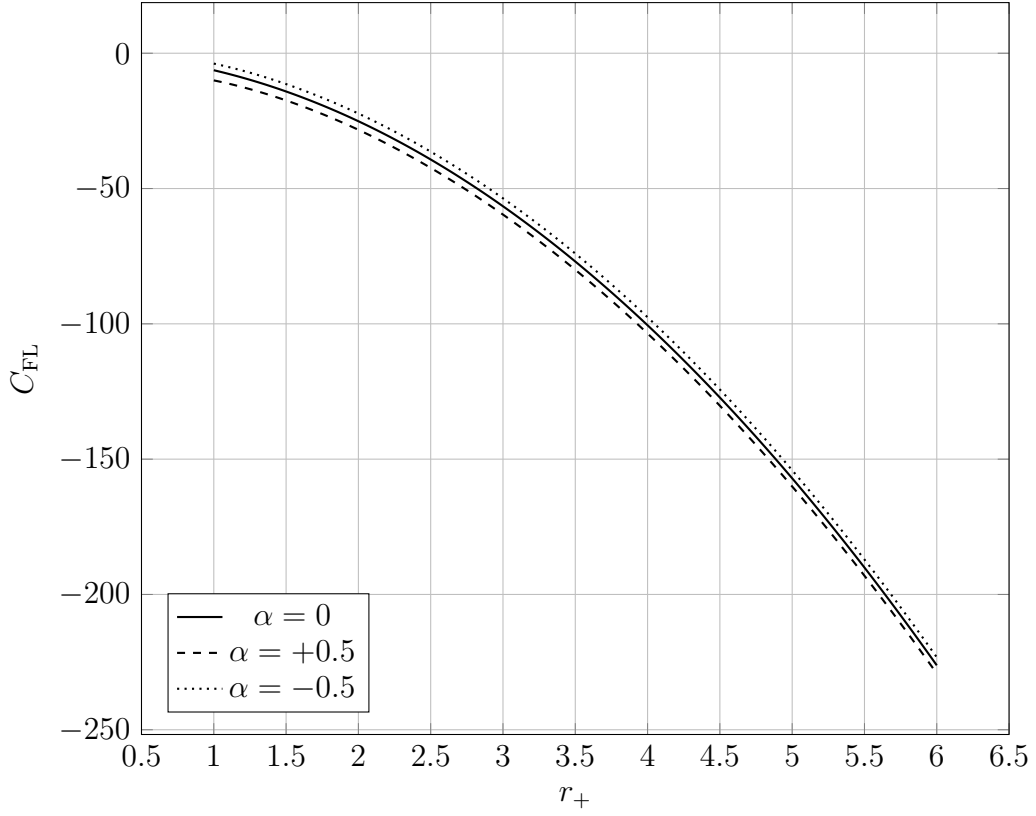

\section{Scalar wave equation and superradiance no-go theorem}\label{sec:scattering}

\subsection{Four-dimensional Klein--Gordon equation}
Consider a minimally coupled scalar field of mass $\mu$:\cite{Wald1993,IyerWald1994,HeisenbergEntropy2022}
\begin{equation}\label{eq:KG-action}
  I_\Phi
  =-\frac12\int\dd^4x\sqrt{-g}
  \left(g^{\mu\nu}\partial_\mu\Phi^*\partial_\nu\Phi
  +\mu^2\Phi^*\Phi\right).
\end{equation}
Variation with respect to $\Phi^*$ gives
\begin{equation}\label{eq:KG-covariant}
  (\Box-\mu^2)\Phi=0,
\end{equation}
where
\begin{equation}\label{eq:box-definition}
  \Box\Phi
  =\frac1{\sqrt{-g}}
  \partial_\mu\left(\sqrt{-g}g^{\mu\nu}\partial_\nu\Phi\right).
\end{equation}
For \eqref{eq:Schwarzschild-line-element},
\begin{equation}
  g_{\mu\nu}
  =\operatorname{diag}\left(-F,F^{-1},r^2,r^2\sin^2\theta\right),
\end{equation}
so
\begin{equation}
  g^{\mu\nu}
  =\operatorname{diag}\left(-F^{-1},F,r^{-2},(r^2\sin^2\theta)^{-1}\right).
\end{equation}
The determinant is
\begin{align}
  g
  &=(-F)(F^{-1})(r^2)(r^2\sin^2\theta)
  \\
  &=-r^4\sin^2\theta,
\end{align}
thus
\begin{equation}
  \sqrt{-g}=r^2\sin\theta.
\end{equation}
Substitution into \eqref{eq:box-definition} gives
\begin{align}
  \Box\Phi
  =&-\frac1F\partial_t^2\Phi
  +\frac1{r^2}\partial_r\left(r^2F\partial_r\Phi\right)
  \nonumber\\
  &+\frac1{r^2\sin\theta}\partial_\theta
  \left(\sin\theta\partial_\theta\Phi\right)
  +\frac1{r^2\sin^2\theta}\partial_\phi^2\Phi.
\end{align}
Define the unit-sphere Laplacian
\begin{equation}
  \Delta_{S^2}
  =\frac1{\sin\theta}\partial_\theta
  (\sin\theta\partial_\theta)
  +\frac1{\sin^2\theta}\partial_\phi^2.
\end{equation}
The wave equation is therefore
\begin{equation}\label{eq:KG-expanded}
  -\frac1F\partial_t^2\Phi
  +\frac1{r^2}\partial_r(r^2F\partial_r\Phi)
  +\frac1{r^2}\Delta_{S^2}\Phi
  -\mu^2\Phi=0.
\end{equation}

\subsection{Separation of variables}
Use
\begin{equation}\label{eq:separation}
  \Phi(t,r,\theta,\phi)
  =e^{-i\omega t}Y_{\ell m}(\theta,\phi)\frac{u_\ell(r)}{r}.
\end{equation}
The angular harmonics satisfy
\begin{equation}
  \Delta_{S^2}Y_{\ell m}
  =-\ell(\ell+1)Y_{\ell m}.
\end{equation}
Also
\begin{equation}
  \partial_t^2e^{-i\omega t}=-\omega^2e^{-i\omega t}.
\end{equation}
After dividing by $e^{-i\omega t}Y_{\ell m}$, \eqref{eq:KG-expanded} becomes
\begin{equation}\label{eq:radial-before-u}
  \frac{\omega^2}{F}\frac{u}{r}
  +\frac1{r^2}\frac{\dd}{\dd r}
  \left[r^2F\frac{\dd}{\dd r}\left(\frac{u}{r}\right)\right]
  -\frac{\ell(\ell+1)}{r^2}\frac{u}{r}
  -\mu^2\frac{u}{r}=0.
\end{equation}
Now
\begin{equation}
  \frac{\dd}{\dd r}\left(\frac{u}{r}\right)
  =\frac{u'}r-\frac{u}{r^2},
\end{equation}
so
\begin{equation}
  r^2F\frac{\dd}{\dd r}\left(\frac{u}{r}\right)
  =rFu'-Fu.
\end{equation}
Differentiate:
\begin{align}
  \frac{\dd}{\dd r}(rFu'-Fu)
  &=(F+rF')u'+rFu''-F'u-Fu'
  \\
  &=rFu''+rF'u'-F'u.
\end{align}
Substitution into \eqref{eq:radial-before-u} and multiplication by $r$ give
\begin{equation}\label{eq:radial-r}
  Fu''+F'u'-\frac{F'}r u
  +\left[
  \frac{\omega^2}{F}
  -\frac{\ell(\ell+1)}{r^2}
  -\mu^2
  \right]u=0.
\end{equation}

\subsection{Tortoise coordinate and effective potential}
Define
\begin{equation}\label{eq:tortoise}
  \frac{\dd r_*}{\dd r}=\frac1F.
\end{equation}
Then
\begin{equation}
  \frac{\dd u}{\dd r_*}=F\frac{\dd u}{\dd r}=Fu',
\end{equation}
and
\begin{align}
  \frac{\dd^2u}{\dd r_*^2}
  &=F\frac{\dd}{\dd r}(Fu')
  \\
  &=F(F'u'+Fu'')
  \\
  &=FF'u'+F^2u''.
  \label{eq:tortoise-second}
\end{align}
Multiply \eqref{eq:radial-r} by $F$:
\begin{align}
  F^2u''+FF'u'
  +\left[
  \omega^2
  -F\left(\frac{F'}r+\frac{\ell(\ell+1)}{r^2}+\mu^2\right)
  \right]u=0.
\end{align}
Using \eqref{eq:tortoise-second},
\begin{equation}\label{eq:Schrodinger-radial}
  \frac{\dd^2u}{\dd r_*^2}
  +\left[\omega^2-V_\ell(r)\right]u=0,
\end{equation}
where
\begin{equation}\label{eq:V-general}
  V_\ell(r)
  =F(r)\left[
  \frac{\ell(\ell+1)}{r^2}
  +\frac{F'(r)}r
  +\mu^2
  \right].
\end{equation}
For Schwarzschild,
\begin{equation}
  F'(r)=\frac{2GM}{r^2},
\end{equation}
so
\begin{equation}\label{eq:V-Schwarzschild}
  V_\ell(r)
  =F(r)\left[
  \frac{\ell(\ell+1)}{r^2}
  +\frac{2GM}{r^3}
  +\mu^2
  \right].
\end{equation}
The potential is real for real $r>r_+$.

\subsection{Conserved Wronskian}
For real $\omega$ and real $V_\ell$, define
\begin{equation}\label{eq:Wronskian}
  W[u,u^*]
  \equiv u^*\frac{\dd u}{\dd r_*}
  -u\frac{\dd u^*}{\dd r_*}.
\end{equation}
Differentiate:
\begin{align}
  \frac{\dd W}{\dd r_*}
  &=\frac{\dd u^*}{\dd r_*}\frac{\dd u}{\dd r_*}
  +u^*\frac{\dd^2u}{\dd r_*^2}
  -\frac{\dd u}{\dd r_*}\frac{\dd u^*}{\dd r_*}
  -u\frac{\dd^2u^*}{\dd r_*^2}
  \\
  &=u^*u''-u(u^*)''.
\end{align}
Using \eqref{eq:Schrodinger-radial},
\begin{equation}
  u''=-(\omega^2-V_\ell)u,
  \qquad
  (u^*)''=-(\omega^2-V_\ell)u^*,
\end{equation}
so
\begin{equation}
  \frac{\dd W}{\dd r_*}
  =-(\omega^2-V_\ell)u^*u
  +(\omega^2-V_\ell)uu^*=0.
\end{equation}
Therefore
\begin{equation}\label{eq:W-constant}
  W=\text{constant}.
\end{equation}

\subsection{Massless reflection--transmission relation}
For $\mu=0$, $V_\ell\to0$ at the horizon and at infinity. Choose a unit-amplitude incident wave from infinity:
\begin{align}
  u&\sim\mathcal T e^{-i\omega r_*},
  &&r_*\to-\infty,
  \label{eq:bc-horizon}\\
  u&\sim e^{-i\omega r_*}+\mathcal R e^{+i\omega r_*},
  &&r_*\to+\infty.
  \label{eq:bc-infinity}
\end{align}
At the horizon,
\begin{equation}
  u_H=\mathcal T e^{-i\omega r_*},
  \qquad
  u_H'=-i\omega\mathcal T e^{-i\omega r_*}.
\end{equation}
Substitution into \eqref{eq:Wronskian} gives
\begin{equation}\label{eq:W-horizon}
  W_H=-2i\omega|\mathcal T|^2.
\end{equation}
At infinity, the incident contribution has Wronskian $-2i\omega$, the reflected contribution has $+2i\omega|\mathcal R|^2$, and the oscillatory cross terms cancel. Hence
\begin{equation}\label{eq:W-infinity}
  W_\infty=-2i\omega(1-|\mathcal R|^2).
\end{equation}
Conservation $W_H=W_\infty$ yields
\begin{equation}
  -2i\omega|\mathcal T|^2
  =-2i\omega(1-|\mathcal R|^2).
\end{equation}
For $\omega>0$, divide by $-2i\omega$:
\begin{equation}\label{eq:RT-relation}
  1-|\mathcal R|^2=|\mathcal T|^2.
\end{equation}
Therefore
\begin{equation}\label{eq:no-superradiance-result}
  |\mathcal R|^2=1-|\mathcal T|^2\leq1.
\end{equation}
There is reflection from the potential barrier, but no amplification.

\subsection{Massive field}
For $\omega>\mu$, the asymptotic wave number is
\begin{equation}
  k_\infty=\sqrt{\omega^2-\mu^2}.
\end{equation}
The infinity boundary condition becomes
\begin{equation}
  u\sim e^{-ik_\infty r_*}
  +\mathcal R e^{+ik_\infty r_*}.
\end{equation}
The horizon wave number remains $\omega$ because $F\mu^2\to0$ at $r=r_+$. Wronskian conservation gives
\begin{equation}
  k_\infty(1-|\mathcal R|^2)
  =\omega|\mathcal T|^2.
\end{equation}
Since $k_\infty>0$ and $\omega>0$,
\begin{equation}
  |\mathcal R|^2
  =1-\frac{\omega}{k_\infty}|\mathcal T|^2
  \leq1.
\end{equation}
For $0<\omega<\mu$, no propagating incident wave exists at spatial infinity.

\subsection{General rotating or charged condition}
For a field with azimuthal number $m$ and electric charge $q$ on a rotating and/or charged black hole, the ingoing horizon mode is governed by the horizon-frame frequency
\begin{equation}\label{eq:omega-tilde}
  \widetilde\omega_H
  =\omega-m\Omega_H-q\Phi_H.
\end{equation}
The horizon Wronskian becomes
\begin{equation}
  W_H=-2i\widetilde\omega_H|\mathcal T|^2.
\end{equation}
For a massless asymptotic wave, conservation gives
\begin{equation}\label{eq:general-RT}
  |\mathcal R|^2
  =1-\frac{\widetilde\omega_H}{\omega}|\mathcal T|^2.
\end{equation}
Amplification requires
\begin{equation}
  |\mathcal R|^2>1,
\end{equation}
which is possible only if
\begin{equation}
  \widetilde\omega_H<0.
\end{equation}
Thus the standard superradiant window is
\begin{equation}\label{eq:superradiant-window}
  0<\omega<m\Omega_H+q\Phi_H
\end{equation}
when the upper bound is positive \cite{Bekenstein1998,Brito2015}.

For the static neutral Schwarzschild background,
\begin{equation}
  \Omega_H=0,
  \qquad
  \Phi_H=0,
  \qquad
  \widetilde\omega_H=\omega>0.
\end{equation}
The interval \eqref{eq:superradiant-window} is empty.

\subsection{Why the logarithmic entropy parameter does not enter the wave equation}
Under \eqref{eq:no-backreaction-assumption},
\begin{equation}
  \frac{\partial F}{\partial\alpha}=0.
\end{equation}
Equation \eqref{eq:V-Schwarzschild} then gives
\begin{align}
  \frac{\partial V_\ell}{\partial\alpha}
  &=\frac{\partial V_\ell}{\partial F}
  \frac{\partial F}{\partial\alpha}
  +\frac{\partial V_\ell}{\partial F'}
  \frac{\partial F'}{\partial\alpha}
  \\
  &=0.
  \label{eq:dV-dalpha}
\end{align}
To see the consequence perturbatively, write
\begin{equation}
  u=u_0+\alpha u_1+\Ord(\alpha^2),
  \qquad
  \mathcal R=\mathcal R_0+\alpha\mathcal R_1+\cdots,
  \qquad
  \mathcal T=\mathcal T_0+\alpha\mathcal T_1+\cdots.
\end{equation}
The wave operator is
\begin{equation}
  \mathcal L_0
  \equiv\frac{\dd^2}{\dd r_*^2}+\omega^2-V_\ell(r),
\end{equation}
with no $\alpha$ dependence. At zeroth order,
\begin{equation}
  \mathcal L_0u_0=0.
\end{equation}
At first order,
\begin{equation}
  \mathcal L_0u_1=0.
\end{equation}
The incident amplitude is fixed to unity independently of $\alpha$, and no first-order source or boundary perturbation is present. The unique first-order correction consistent with those homogeneous conditions is
\begin{equation}
  u_1=0,
  \qquad
  \mathcal R_1=0,
  \qquad
  \mathcal T_1=0.
\end{equation}
Therefore
\begin{equation}\label{eq:no-alpha-scattering-final}
  \frac{\partial\mathcal R}{\partial\alpha}=0,
  \qquad
  \frac{\partial\mathcal T}{\partial\alpha}=0
\end{equation}
within the assumptions of this paper.

A nonzero correction would require an explicit semiclassical equation producing
\begin{equation}
  F(r;\alpha)=F_0(r)+\alpha F_1(r)+\Ord(\alpha^2),
\end{equation}
or an explicit correction to $\Omega_H$, $\Phi_H$, or the matter action. None of these quantities can be derived from the entropy logarithm alone.

\section{Discussion}\label{sec:discussion}
The detailed calculation establishes four points.

First, an $f(\Qsc)$ background is a metric--connection pair. The connection can vanish in affine coordinates, but a nonlinear transformation to a standard spherical chart generally produces nonzero inertial components. The explicit pair in \Cref{sec:schwarzschild} is flat and torsion-free in the affine sector and solves the STEGR metric equation.

Second, the area entropy is not inserted blindly into an arbitrary modified-gravity solution. It is justified on the STEGR branch by the extended Wald construction and independently reconstructed from the classical first law.

Third, the three quantities $M$, $T$, and $S$ cannot all be modified independently while retaining $\dd M=T\dd S$. Adding $\alpha\ln S_0$ while keeping the classical mass and geometric temperature produces the explicit mismatch \eqref{eq:first-law-mismatch}. The effective temperature \eqref{eq:TFL-final} is one consistent thermodynamic convention, but it is not the geometric Hawking temperature of a corrected spacetime.

Fourth, superradiance is a flux statement. The sign capable of producing amplification is the sign of the horizon-frame frequency $\widetilde\omega_H$. A static neutral black hole has no rotational or electromagnetic chemical potential, so $\widetilde\omega_H=\omega>0$. A logarithmic entropy term supplies neither a negative horizon-frame frequency nor a modified wave operator.

\section{Conclusion}\label{sec:conclusion}
We have supplied a step-by-step, reproducible benchmark for logarithmic entropy corrections in symmetric teleparallel gravity. The main results are:
\begin{enumerate}
  \item The nonmetricity scalar, superpotential, $f_{\Qsc}$, $f_{\Qsc\Qsc}$, metric equation, connection equation, pure-gauge connection, and STEGR boundary-term identity have been displayed explicitly.
  \item Solving the vacuum Einstein-form equations gives
  \begin{equation*}
    F(r)=1-\frac{2GM}{r},
    \qquad
    r_+=2GM,
    \qquad
    M=\frac{r_+}{2G}.
  \end{equation*}
  \item The surface gravity and Euclidean regularity calculations both give
  \begin{equation*}
    T_\kappa=\frac1{4\pi r_+}.
  \end{equation*}
  \item The leading entropy is
  \begin{equation*}
    S_0=\frac{A_H}{4G}=\frac{\pi r_+^2}{G}.
  \end{equation*}
  \item With
  \begin{equation*}
    S_{\mathrm{corr}}=S_0+\alpha\ln(S_0/S_\star),
  \end{equation*}
  the fixed-mass first-law temperature is
  \begin{equation*}
    T_{\mathrm{FL}}
    =\frac{r_+}{4(\pi r_+^2+\alpha G)},
  \end{equation*}
  while $T_\kappa$ remains unchanged without backreaction.
  \item The effective heat capacity is
  \begin{equation*}
    C_{\mathrm{FL}}
    =\frac{2(\pi r_+^2+\alpha G)^2}
    {G(\alpha G-\pi r_+^2)}.
  \end{equation*}
  Its formal pole occurs at $S_0=\alpha$, outside the regime $|\alpha|/S_0\ll1$.
  \item The scalar radial equation has the real potential
  \begin{equation*}
    V_\ell
    =F\left[\frac{\ell(\ell+1)}{r^2}
    +\frac{2GM}{r^3}+\mu^2\right],
  \end{equation*}
  and Wronskian conservation yields $|\mathcal R|^2\leq1$.
  \item Since $\partial_\alpha F=0$, one has $\partial_\alpha V_\ell=0$ and no first-order correction to the reflection or transmission amplitudes.
\end{enumerate}

A genuine modified-gravity calculation of superradiant corrections must begin with an explicit rotating or charged $f(\Qsc)$ metric--connection solution and a branch-specific entropy. A semiclassical effective action must then determine the corrected geometry and horizon potentials. The logarithmic entropy term alone is insufficient.

\bigskip
\noindent\textbf{Data availability.} No external numerical data are used. Every table entry and curve follows directly from the analytic equations displayed in the text.

\bigskip
\noindent\textbf{Conflict of interest.} The author declares no conflict of interest.

\end{document}